\documentclass[runningheads,a4paper]{llncs}

\usepackage{amssymb}
\usepackage{graphicx}
\usepackage{amsmath}
\usepackage{mathtools}
\usepackage{subfig}
\usepackage{url}
\usepackage{etoolbox}
\patchcmd{\footnotemark}{\stepcounter{footnote}}{\refstepcounter{footnote}}{}{}
\urldef{\mailsa}\path|{piyush.bansal, romil.bansal}@research.iiit.ac.in,|    
\urldef{\mailsb}\path|vv@iiit.ac.in|    
\newcommand{\keywords}[1]{\par\addvspace\baselineskip
\noindent\keywordname\enspace\ignorespaces#1}

\begin{document}

\mainmatter  

\title{Towards Deep Semantic Analysis of Hashtags}

\titlerunning{Towards Deep Semantic Analysis of Hashtags}

%
%
\author{Piyush Bansal%
\and Romil Bansal\and Vasudeva Varma}
\authorrunning{Piyush Bansal%
\and Romil Bansal\and Vasudeva Varma}

\institute{International Institute of Information Technology\\
Hyderabad, Telangana, India\\
\mailsa\\
\mailsb\\}

%
%

\toctitle{Lecture Notes in Computer Science}
\tocauthor{Piyush Bansal, Romil Bansal and Vasudeva Varma}
\maketitle

\begin{abstract}
Hashtags are semantico-syntactic constructs used across various social networking and microblogging platforms to enable users to start a topic specific discussion or classify a post into a desired category. Segmenting and linking the entities present within the hashtags could therefore help in better understanding and extraction of information shared across the social media. However, due to lack of space delimiters in the hashtags (e.g \emph{\#nsavssnowden}), the segmentation of hashtags into constituent entities (\emph{``NSA"} and \emph{``Edward Snowden"} in this case) is not a  trivial task. Most of the current state-of-the-art social media analytics systems like Sentiment Analysis and Entity Linking tend to either ignore hashtags, or treat them as a single word. 
In this paper, we present a context aware approach to segment and link entities in the hashtags to a knowledge base (KB) entry, based on the context within the tweet.
Our approach segments and links the entities in hashtags such that the coherence between hashtag semantics and the tweet is maximized. To the best of our knowledge, no existing study addresses the issue of linking entities in hashtags for extracting semantic information. 
We evaluate our method on two different datasets, and demonstrate the effectiveness of our technique in improving the overall entity linking in tweets via additional semantic information provided by segmenting and linking entities in a hashtag.

\keywords{Hashtag Segmentation, Entity Linking, Entity Disambiguation, Information Extraction}
\end{abstract}

\section{Introduction}

Microblogging and Social Networking websites like \emph{Twitter, Google+, Facebook} and \emph{Instagram} are becoming 
increasingly popular with more than 400 million posts each day. 
This huge collection of posts on the social media makes it an important source for 
gathering real-time news and event information. Microblog posts are often tagged with an unspaced phrase,      prefixed with the sign ``\#" known as a hashtag. 
14\% of English tweets are tagged with at least 1 hashtag with 1.4 hashtags per tweet~\cite{wouter2011}. 
Hashtags make it possible to categorize and track a 
microblog post among millions of other posts. 
Semantic analysis of hashtags could therefore help us in understanding and extracting important information from microblog posts.

In English, and many other Latin alphabet based languages, 
the inherent structure of the language imposes an assumption, under which the space character is a good approximation of 
word delimiter. However, hashtags violate such an assumption making it difficult to analyse them.
In this paper, we analyse the problem of extracting semantics in hashtags by segmenting and linking entities within hashtags. 
For example, given a hashtag like ``\#NSAvsSnowden" occurring inside a tweet, 
we develop a system that not only segments the hashtag into ``NSA vs Snowden", 
but also tells that ``NSA" refers to ``National Security Agency" and ``Snowden" refers to ``Edward Snowden". 
Such a system has numerous applications in the areas of Sentiment Analysis, Opinion Mining, Event Detection 
and improving quality of search results on Social Networks, as these systems can leverage additional semantic information provided by the hashtags present within the tweets. 
Our system takes a hashtag and the corresponding tweet text as input and returns the segmented hashtag along with 
Wikipedia pages corresponding to the entities in the hashtag. 
To the best of our knowledge, the proposed system is the first to focus on extracting semantic knowledge from hashtags 
by segmenting them into constituent entities.

\section{Related Work}
\label{sec:related}
The problem of word segmentation has been studied in various contexts in the past. 
A lot of work has been done on Chinese word segmentation.  Huang et al.~\cite{huang07} showed that character based tagging approach 
outperforms other word based segmentation approaches for Chinese word segmentation. 
English URL segmentation has also been explored by various researchers in the past~\cite{wang2011}\cite{kan05_cikm}\cite{srinivasan12_cikm}. 
All such systems explored length specific features to segment the URLs into constituent 
chunks\footnote{The term ``chunk" here and henceforth refers to each of the segments \(s_{i}\) in a segmentation \(S\) = \(s_{1}, s_{2}, ... s_{i}, ... s_{n}\). 
For example, in case of the hashtag \#NSAvsSnowden, one of the possible segmentations (\(S\)) is NSA, vs, Snowden. Here, the \emph{words} - ``NSA", ``vs" and ``Snowden" are being referred to as chunks.}. 
Although a given hashtag can be segmented into various possible segments, all of which are plausible, the ``correct" segmentation depends on the tweet context. 
For example, consider a hashtag `\emph{notacon}'. It can be segmented into chunks `not, a, con' or `nota, con' based on the tweet context.
The proposed system focuses on hashtag segmentation while being context aware.
Along with unigram, bigram and domain specific features, content in the tweet text is also considered for segmenting and linking the entities 
within a  
hashtag.

Entity linking in microposts has also been studied by various researchers recently. 
Various features like commonness, relatedness, popularity and recentness have been used for detecting and linking the entities 
in the microposts~\cite{meij}\cite{tagme}\cite{bansal}. Although semantic analysis of microposts has been studied vastly, hashtags are either ignored or treated as a single word. 
In this work, we analyse hashtags by linking entities in the hashtags to the corresponding Wikipedia page.

\section{System Architecture}

In this section, we present an overview of our system.
We also describe the features extracted, followed by a discussion on training and learning procedures in Section~\ref{sec:trainingprocedures}.

\begin{figure}
\centering
\includegraphics[height=4.5cm]{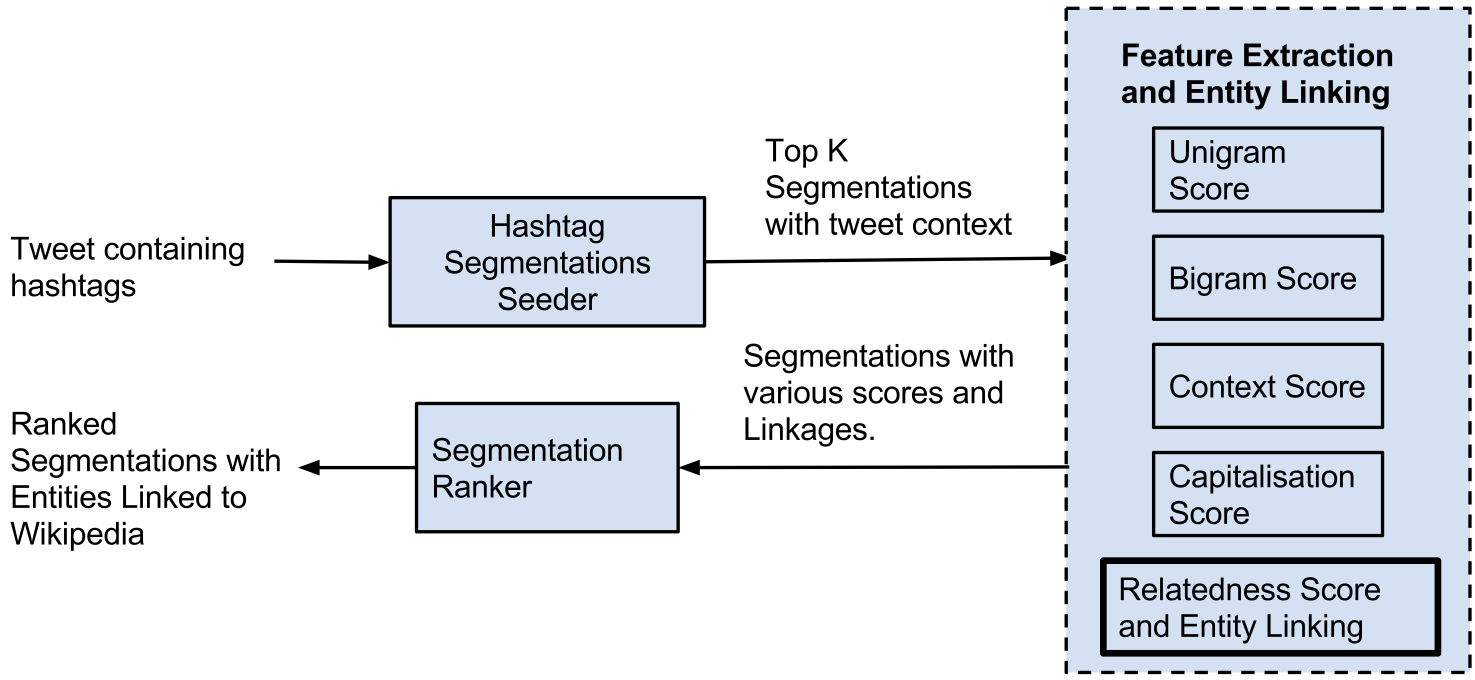}
\caption{Schematic Diagram of the Overall System.}
\label{fig:example}
\end{figure}

As illustrated in Fig.~\ref{fig:example}, the proposed system has 3 major components - 1) \emph{Hashtag Segmentations Seeder}, 2) \emph{Feature Extraction and Entity Linking module}, and 3) \emph{Segmentation Ranker}. In the following sections, we describe each component in detail.

\subsection{Hashtag Segmentations Seeder }

Hashtag Segmentations Seeder is responsible for generating a list of possible segmentations of a given hashtag. We propose \emph{Variable Length Sliding Window technique} for generating a set of highly probable hashtag segmentations for the given hashtag in the first step. 

The \emph{Variable Length Sliding Window technique} is based on an assumption that for a given hashtag ``\textit{{\#AXB}}'', if A and B are valid semantic units (a single word or a collection of words concatenated together without a space), it is reasonable to hypothesize that X is also a valid semantic unit. For example, in the hashtag ``\textit{{\#followUCBleague}}'', since, `\textit{{follow}}' and `\textit{{league}}' are well known dictionary words, and collectively this hashtag has some semantic meaning associated with it as it has occurred in a tweet, it is reasonable to assume that `\textit{{UCB}}' is also a valid semantic unit with some meaning associated with it.
The length of the sliding window(\(X\)) is varied from \texttt{MIN\_LEN} to \texttt{MAX\_LEN} with each iteration, and the window is slid over the hashtag. \(O(n^2\)) triplets of the form (A, X, B) are generated using the sliding window technique, where n is the length of the hashtag, X is the part of the hashtag lying within the window and A and B are the parts of the hashtag (of length $\geq$ 0) that lie on the left and right of the window respectively.

Each segment $A$ and $B$ of the triplet (A, X, B) is assigned a score according to the classically known Dynamic Programming based algorithm for Word Segmentation~\cite{norvig03}, hereby referred to as \(ViterbiWordSeg\).\\
\(ViterbiWordSeg\) takes a string as input and returns the best possible segmentation \(BestSeg\) (ordered collection of chunks) for that string. The score assigned to the segmentation by \(ViterbiWordSeg\) is the sum of log of probability scores of the segmented chunks based on the unigram language model.
\begin{equation}
ViterbiWordSegScore(S) = \sum_{s_i\in BestSeg(S)}log(P_{Unigram}(s_i))
\end{equation}

\begin{figure}
\centering
\includegraphics[width=6cm]{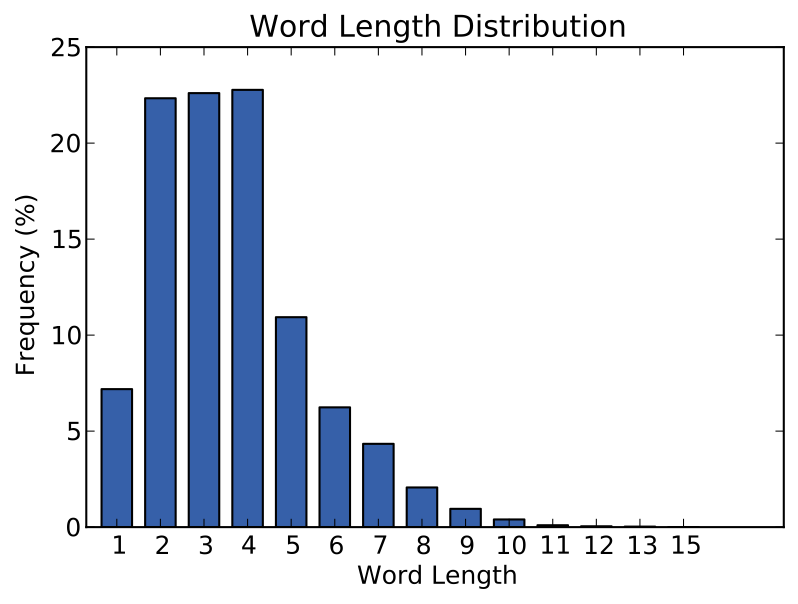}
\caption{Word Length vs. Frequency Percentage graph for 50M tweets.}
\label{freq}
\end{figure}

We used Microsoft Web N-Gram Services\footnote{Microsoft Web N-Gram Services~\url{http://research.microsoft.com/en-us/collaboration/focus/cs/web-ngram.aspx}} for computing the unigram probability scores. The aforementioned corpus contains data from the web, and hence various acronyms and slang words occur in it. This holds critical importance in the context of our task. Next, for each triplet of the form (A, X, B), we compute the Sliding Window score as follows.
\begin{equation}
\begin{multlined}
\label{eq:oov}
Score_{SlidingWindow}(A, X, B) = ViterbiWordSegScore(A) + \\
constant * \log_{10}(UnigramProb(X)) * WordLenProb(len(X)) + \\
ViterbiWordSegScore(B)
\end{multlined}
\end{equation}

where \(WordLenProb(x)\) is the \(Ordinate\) value at x in Figure~\ref{freq} and the  \texttt{constant} is set by experimentation.

Also, for each triplet (A, X, B), the final segmentation,
\(Seg(A, X, B)\) is the ordered collection of chunks (\(BestSeg(A)\), X, \(BestSeg(B)\)), where \(BestSeg(A)\) and \(BestSeg(B)\) refer to the best segmentation (ordered collection of chunks) returned by \(ViterbiWordSeg(A)\) and \(ViterbiWordSeg(B)\) respectively.

To find the suitable value of \texttt{MIN\_LEN} and \texttt{MAX\_LEN}, we plot the percentage of frequency vs. word length graph using 50 million tweets\footnote{The dataset is available at
\url{http://demeter.inf.ed.ac.uk/cross/docs/fsd_corpus.tar.gz}}. Figure~\ref{freq} shows the plot obtained.

It is observed that 79\% of the tweet words are between length 2 to 6. Hence, we set \texttt{MIN\_LEN} and \texttt{MAX\_LEN} as 2 and 6 respectively.

The major benefit of this technique is that we are able to handle named entities and out of vocabulary (OOV) words. This is achieved by assigning score as a function of \(WordLenProb\) and smoothed backoff unigram probability (Equation ~\ref{eq:oov}) for words within the window.

Now that we have a list of \(O(n^2\)) segmentations and their corresponding \(Score_{SlidingWindow}\), we pick the top \emph{k} segmentations for each hashtag on the basis of this score. We set \(k = 20\), as precision at 20 (\(P@20\)) comes out to be around 98\%. This establishes that the subset of segmentations we seed, which is of \(O(n^2)\), indeed contains highly probable segmentations out of a total possible \(2^{n-1}\) segmentations\footnote{For a string made up of \(n\) characters, we need to decide where to put the spaces so that we can get a sequence of valid words. There are \(n-1\) positions where a space can be placed, and each position may or may not have a space. Hence there are \(2^{n-1}\) segmentations.}.

\subsection{Feature Extraction and Entity Linking }

\label{sec:featureextraction}
This component of the system is responsible for two major tasks, feature extraction from each of the seeded segmentations, and entity linking on the segmentations. The features, as also shown in the System Diagram, are 1) Unigram Score, 2) Bigram Score, 3) Context Score, 4) Capitalisation Score, and 5) Relatedness Score. The first feature, Unigram Score, is essentially the \(ViterbiWordSegScore\) computed in the previous step. In the following sections, we describe the rest of the features.

\subsubsection{Bigram Score: }

For each of the segmentations seeded by the \emph{Variable Length Sliding Window Technique}, a bigram based score using the Microsoft Web N-Gram Services is computed. It is possible for a hashtag to have two perfectly valid segmentations. Consider the hashtag \textit{\textbf{\#Homesandgardens}}. Now this hashtag can be split as ``Homes and gardens'' which seems more probable to occur in a given context than ``Home sand gardens''. Bigram based scoring helps to rank such segmentations, so that higher scores are awarded to the more semantically ``appealing'' segmentations. The bigram language model would score one of the above segmentations - ``Homes and gardens'' as
\begin{equation}
\begin{multlined}
P(Homes,and,gardens) \approx  \\
P(Homes|<s>) * P(and|Homes) * \\
P(gardens|and) * P(</s>|gardens)
\end{multlined}
\end{equation}

\subsubsection{Context Score: } 

Context based score is an important feature. This is responsible for bubbling up of the segmentations with maximum contextual similarity with the tweet content. Using the CMU TweetNLP toolkit~\cite{gimpel11}, words having POS tags like verb, noun and adjective are extracted both from the candidate segmentation of the hashtag and the tweet context, i.e. the text of the tweet other than the hashtag. Next, Wu Palmer similarity from Wordnet~\cite{miller98} is used on these two sets of words to find how similar a candidate segmentation is to the tweet context. These scores are normalized from 0 to 1.

\subsubsection{Capitalisation Score: }

Hashtags are of varied nature. Some hashtags have a camelcase-like capitalisation pattern as in \textit{\textbf{\#HomesAndGardens}}, while others have everything in lowercase or uppercase characters like \textit{\textbf{\#homesandgardens}}. However, we can easily see that camelcase conveys more information as it helps segment the hashtag into ``Homes and gardens'' and not ``Home sAnd Gardens''. Capitalisation score helps us to capture the information conveyed by capitalisation patterns within the hashtags. We use the following two rules.
For a hashtag, 
\begin{itemize}

\item If a set of characters occuring together are in capitals as in \textit{\textbf{\#follow\underline{UCB}league}}, they are considered to be a part of an ``assumed cluster'' (``UCB" in this case).
\item If it has a few capital letters separated by a group of lower case letters as in \textit{\textbf{\#SomethingGood}}, we assume the capital letters are delimiters and hence derive a few assumed clusters from the input hashtag.

\end{itemize}

We calculate the capitalisation score for a given segmentation \(S\) containing chunks \(s_{1}, s_{2} ... s_{i} .. s_{n} \) as
\begin{equation}
Score_{Cap} = \sum\limits_{i=1}^n assumedClusterNotIntact(s_{i})  
\end{equation}
where \(assumedClusterNotIntact(s_{i})\) returns \(1\), if \(s_{i}\) fails to keep an assumed cluster intact, and \(0\) otherwise.

\subsubsection{Relatedness Score: }

Relatedness score measures the coherence between the tweet context and the hashtag segmentation. This score is computed on the basis of semantic relatedness between the entities present within the segmented hashtag and the tweet context.

We calculated the relatedness between all the possible mentions 
in the segmented hashtag ($M_H$) to all other possible mentions in the tweet context ($M_T$). For computing relatedness between the two entities, we used the Wikipedia-based relatedness function as proposed by 
Milne and Witten~\cite{milne}. 

Relatedness between two Wikipedia pages $p_a$ and $p_b$ is defined as follows:
\begin{equation}
rel(p_a,p_b) = 1 - \delta
\end{equation}
where,
\begin{equation}
\delta = \frac{log(max(|in(p_a), in(p_b)|)) - log(|in(p_a)\cap in(p_b)|)}{log(W) - log(min(|in(p_a),in(p_b)|))}
\end{equation}
$in(p_a)$ is the set of Wikipedia pages pointing to page $p_a$ and $W$ is the total number of pages in Wikipedia.\\
The overall vote given to a candidate page $p_a$ for a given mention $a$ by a mention $b$ is defined as
\begin{equation}
vote_b(p_a) = \frac{\sum_{p_b \in Pg(b)} rel(p_b,p_a). Pr(p_b|b)}{|Pg(b)|}
\end{equation}
where $Pg(b)$ are all possible candidate pages for the mention $b$ and $Pr(p_b|b)$ is the prior probability of b linking to a page $p_b$.\\
The total relatedness score given to a candidate page  $p_a$ for a given mention $a$ is the sum of votes from all other mentions in the tweet context ($M_T$).
\begin{equation}
rel_a(p_a) = \sum_{b \in M_T} vote_b(p_a)
\end{equation}

Now the overall relatedness score for a given hashtag segmentation, $h$ is 
\begin{equation}
score_{h} = \frac{\sum_{m \in M_H} rel_m(p_a).Pr(p_a|m)}{|M_H|}  
\end{equation}

The detected page $p_a$ for a given mention in the segmented hashtag is the Wikipedia page with the highest $rel_a(p_a)$. Since not all the entities are meaningful, we prune the entities with very low $rel_a(p_a)$ scores. In our case, the threshold is set to 0.1. This disambiguation function is considered as state-of-the-art and has also been adopted by various other systems~\cite{tagme}\cite{zhao}.
The relatedness score, $score_{h}$ is used as a feature for hashtag segmentation. The entities in the segmented hashtag are returned along with the score for further improving the hashtag semantics. 

\subsection{Segmentation Ranker}

This component of the system is responsible for ranking the various probable segmentations seeded by the \emph{Hashtag Segmentations Seeder Module}. We generated five features for each segmentation using \emph{Feature Extraction and Entity Linking Module} in the previous step. These scores are combined by modelling the problem as a regression problem, and the combined score is referred to as \(Score_{Regression}\). The segmentations are ranked using \(Score_{Regression}\). In the end, the \emph{Segmentation Ranker} outputs a ranked list of segmentations along with the entity linkings.

In the next section, we discuss the regression and training procedures in greater detail.

\section{Training Procedure}
\label{sec:trainingprocedures}
For the task of training the model, we consider the \(Score_{Regression}\) of all correct segmentations to be \(1\) and all incorrect segmentations as \(0\). Our feature vector comprises of five different scores calculated in Section~\ref{sec:featureextraction}. We use linear regression with elastic net regularisation~\cite{elasticnet}. This allows us to learn a model that is trained with both L1 and L2 prior as regularizer. It also helps us take care of the situation when some of the features might be correlated to one another. Here, \(\rho\) controls the convex combination of L1 and L2.

The Objective Function we try to minimize is
\begin{equation}
\underset{w}{min\,} { \frac{1}{2n_{samples}} ||X w - y||_2 ^ 2 + \alpha \rho ||w||_1 + \frac{\alpha(1-\rho)}{2} ||w||_2 ^ 2}
\end{equation}
where \(X\), \(y\) and \( w\) are Model Matrix, Response Vector, and Coefficient Matrix respectively. 
The parameters \emph{alpha}\((\alpha)\) and \emph{rho}\((\rho)\) are set by cross validation.

\section{Experiments and Results}

In this section we describe the datasets used for evaluation, and establish the effectiveness of our technique by comparing our results to a well known end-to-end Entity Linking system, TAGME~\cite{tagme}, which works on short texts, including tweets.

\subsection{Evaluation Metrics and Datasets}

This section is divided into two parts. First, we explain the evaluation metrics in the context of our experiments, and later, we discuss the datasets used for evaluation.

\subsubsection{Evaluation Metrics}
\label{metrics}
We evaluated our system on two different metrics. Firstly, the system is evaluated based on its performance in the segmentation task. As the system returns a list of top-k hashtag segmentations for a given hashtag, we evaluated the precision at n (P@n) scores for the hashtag segmentation task. We also compared our P@1 score with Word Breaker\footnote{http://web-ngram.research.microsoft.com/info/break.html}, which does the task of word segmentation. Secondly, the system is also evaluated on the basis of its entity linking performance on the hashtags. We computed Precision, Recall and F-Measure scores for the entities linked in the top ranked hashtag. For Entity Linking task, we used the same notions of Precision, Recall, and F-Measure as proposed by Marco et al.~\cite{benchmarking}. We compared our system with the state-of-the-art TAGME system.

We show that adding semantic information extracted from the hashtags leads to an improvement in the overall tweet entity linking. For this, we performed a comparative study on the output of the TAGME system when a tweet is given with un-segmented hashtag vs. when it is given with segmented and entity-linked hashtag\footnote{For the segmented and entity-linked case, the linked entities in a hashtag were replaced with the corresponding Wikipedia page titles.}. The case when un-segmented hashtag is fed to TAGME is considered as a baseline to show how much improvement can be attributed to our method of enriching the tweet with additional semantic information mined by segmenting and linking entities in a hashtag.

\subsubsection{Datasets}

The lack of availability of a public dataset that suits our task has been a major challenge. To the best of our knowledge, no publicly available dataset contains tweets along with hashtags, and the segmentation of hashtag into constituent entities appropriately linked to a Knowledge Base. So, we approached this problem from two angles - 1) Manually Annotated Dataset Generation (where dataset is made public), 2) Synthetically generated Dataset. The datasets are described in detail below.

\begin{table}[t]
\centering
\captionsetup{justification=centering}
\subfloat[Comparative Accuracies for Hashtags Entity Linking task.]{%
\begin{tabular}{|l|c|c|c|}
\hline
    & Precision & Recall & F Score \\
\hline
TAGME   & 0.441  & 0.383   & 0.410\\
(Baseline)&&&\\
\hline
Our System& 0.711  & 0.841  & 0.771 \\
\hline
\end{tabular}}%
\qquad
\subfloat[Comparative Accuracies for Overall Tweet Entity Linking task.]{%
\begin{tabular}{|l|c|c|c|}
\hline
    & Precision & Recall & F Score\\
\hline
TAGME& 0.63  & 0.69  & 0.658 \\
(Baseline)&&&\\
\hline
Our System& 0.732  & 0.91  &0.811\\
+ TAGME&&&\\
\hline
\end{tabular}}%
\qquad
\subfloat[Various \(P@n\) for Hashtag Segmentation task.]{%
\begin{tabular}{|c|c|c|c|c|c|c|}
\hline
n & 1 & 2 & 3 & 5 & 10 & 20\\
\hline
\(P@n\) &0.914   &0.952   &0.962   &0.970   &0.974   &0.978\\
\hline
\end{tabular}}%
\newline
\newline
\caption[something goes here]{Comparative Accuracies on the Microposts NEEL Dataset.\footnotemark\label{fnm:1}}
\label{table:neel}
\vspace{-20 px}
\end{table}

\footnotetext{``TAGME (Baseline)" refers to the baseline evaluation where we give an unsegmented hashtag to TAGME to annotate. ``Our System + TAGME" refers to the evaluation, where we first do segmentation and entity linking on hashtags using our system, and then feed them to TAGME to annotate either just the hashtag (Table a) or the full tweet (Table b). This is also discussed under ``Evaluation Metrics"in subsection ~\ref{metrics}.}

\paragraph{1. \underline{Microposts NEEL Dataset}: }

The Microposts NEEL Dataset~\cite{microposts} contains over 3.5k tweets collected over a period from 15th July 2011 to 15th August 2011, and is rich in event-annotated tweets. This dataset contains Entities, and the corresponding linkages to DBPedia. The problem however, is that this dataset does not contain the segmentation of hashtags. We generate synthetic hashtags by taking tweets, and combining random number of consecutive words with each entity present within them. The remaining portion of the tweet that does not get combined is considered to be the tweet context. If no entity is present within the tweet, random words are combined to form the hashtag. This solves the problem of requiring human intervention to segment and link hashtags, since now we already know the segmentation as well as the entities present within the hashtag. 

Our system achieved an accuracy (\(P@1\)) of 91.4\% in segmenting the hashtag correctly. The accuracy of Word Breaker in this case was 80.2\%. This, however, can be attributed to a major difference between our system and Word Breaker. Word Breaker is not context aware. It just takes an unspaced string, and tries to break it into words. Our method takes into account the relatedness between the entities in a hashtag and the rest of the tweet content. Also, various other hashtag specific features like Capitalisation Score play an important part in improving the accuracy.

The comparative results of Entity Linking (in hashtags and overall), as well as \(P@n\) at various values of \(n\) for segmentation task are contained in Table~\ref{table:neel}. All the values are calculated by k-fold Cross-validation with k=5.

\begin{table}[t]
\centering
\captionsetup{justification=centering}
\subfloat[Comparative Accuracies for the Hashtag Entity Linking task.]{%
\begin{tabular}{|l|c|c|c|}
\hline
    & Precision & Recall & F Score \\
\hline
TAGME   &  0.398  &  0.465  &  0.429\\
 (Baseline)&&&\\
\hline
 Our System&  0.731  &  0.921  & 0.815\\
\hline
\end{tabular}}%
\qquad
\subfloat[Comparative Accuracies for the Overall Tweet Entity Linking task.]{%
\begin{tabular}{|l|c|c|c|}
\hline
    & Precision & Recall & F Score\\
\hline
TAGME   & 0.647  & 0.732  & 0.687 \\
(Baseline)&&&\\
\hline
Our System& 0.748  & 0.943  & 0.834\\
+ TAGME &&&\\
\hline
\end{tabular}}%
\qquad
\subfloat[Various \(P@n\) for Hashtag Segmentation task.]{%
\begin{tabular}{|c|c|c|c|c|c|c|}
\hline
n & 1 & 2 & 3 & 5 & 10 & 20\\
\hline
\(P@n\) &0.873   &0.917   &0.943   &0.958   &0.965    &0.967\\
\hline
\end{tabular}}%
\newline
\newline
\caption[LOL]{Comparative Accuracies on the Manually Annotated Stanford Sentiment Analysis Dataset.}
\label{table:stanford}
\vspace{-20 px}
\end{table}

\paragraph{2. \underline{Manually Annotated Stanford Sentiment Analysis Dataset}: }

To overcome the limitation that a synthetically generated hashtag might not actually be equivalent to a real world hashtag, we sampled around 1.2k tweets randomly from the Stanford Sentiment Analysis Dataset\footnote{http://cs.stanford.edu/people/alecmgo/trainingandtestdata.zip}, all of which contained one or more hashtags in them. After this, we generated around 20 possible segmentations for each hashtag by passing the hashtag and tweet from \emph{Segmentations Seeder Module}. In the end we had around 21k rows which were given to 3 human annotators to annotate as 0 or 1 depending on whether or not a given segmentation is correct (for a given hashtag) according to their judgement. 

Determining the ``correct" segmentation for a given hashtag is particularly challenging, as there may be many answers that are equally plausible. It has been long established that there exist style disagreements among various editorial content (``Homepage" vs ``Home page"). There are also various new words that come into existence like ``TweetDeck" which are brand or product names. So, our annotation guidelines in case of Stanford Sentiment Analysis Dataset allow for annotators to mark multiple segmentations as correct. 

The rows were labelled 0, if at least 2 annotators out of 3 agreed on the label 0, similarly the rows were labelled 1, if at least 2 out of 3 annotators agreed on the label 1. The labels are essentially \(Score_{Regression}\) as described in Section~\ref{sec:trainingprocedures}. The value of \emph{Fleiss' Kappa} (\(\kappa\)), which is a measure of inter annotator agreement, comes out to be \(0.89\), showing a good agreement between annotators. This dataset is made public to ease future research in this area\footnote{Dataset:  \url{http://bit.ly/HashtagData}}.

Our system achieved a precision (\(P@1\)) of 87.3\% in segmenting the hashtags correctly. The \(P@1\) score of Word Breaker in this case was 78.9\%. The difference in performance can again be attributed to same reasons as in the case of NEEL Dataset. 

The comparative results of Entity Linking (in hashtags and overall), as well as \(P@n\) at various values of \(n\) for the task of segmentation are contained in  Table~\ref{table:stanford}. All the values are calculated by k-fold Cross-validation with k=5.

\vspace{-12 px}
\subsubsection{Results}

We demonstrate the effectiveness of our technique by evaluating on two different datasets. We also show how overall Entity Linking in tweets was improved, when our system was used to segment the hashtag and link the entities in the hashtag. We achieved an improvement of 36.1\% F-Measure in extracting semantics from hashtags over the baseline in case of NEEL Dataset. We further show that extracting semantics led to overall increase in Entity Linking of tweet. In case of NEEL Dataset, we achieved an improvement of 15.3\% F-Measure over baseline in overall tweet Entity Linking task as can be seen in Table~\ref{table:neel}. Similar results were obtained for the Annotated Stanford Sentiment Analysis Dataset as well, as shown in Table~\ref{table:stanford}.
Further, we measured the effectiveness of each feature in ranking the hashtag segmentations. The results are summarized in Table~\ref{table:features}.
\begin{table}[t]
\centering
\begin{tabular}{|ll|c|c|}
\hline
& {\textbf{Added Feature}} & {\textbf{P@1}} & $\Delta$\\
\hline
&Unigram & 0.834 & NA \\
\hline
+& Bigram & 0.846 & +1.2\%  \\
\hline
+& Context & 0.855 & +0.9\% \\
\hline
+& Capitalisation & 0.862 & +0.7\% \\
\hline
+& Relatedness & 0.873 & +1.1\%\\
\hline
\end{tabular}
\vspace{5 mm}
\caption{Importance of each feature}
\label{table:features}
\vspace{-20 px}
\end{table}

\section{Conclusions}

We have presented a context aware method to segment a hashtag, and link its constituent entities to a Knowledge Base (KB). An ensemble of various syntactic, as well as semantic features is used to learn a regression model that returns a ranked list of probable segmentations. This allows us to handle cases where multiple segmentations are acceptable (due to lack of context in cases, where tweets are extremely short) for the same hashtag, e.g. \textit{\textbf{\#Homesandgardens}}.

The proposed method of extracting more semantic information from hashtags can be beneficial to numerous tasks including, but not limited to sentiment analysis, improving search on social networks and microblogs, topic detection etc.

\end{document}